# Timing Calibration of the APOLLO Experiment

James B. R. Battat, Louisa Ruixue Huang, Else Schlerman
*Wellesley College, Department of Physics, Wellesley, MA 02481*

Thomas W. Murphy, Jr., Nick R. Colmenares, Rodney Davis
*University of California, San Diego, Department of Physics, La Jolla, CA 92023*

**Abstract**
The Apache Point Observatory Lunar Laser-ranging Operation (APOLLO) began millimeter-precision ranging to the Moon in 2006. Until now, a comprehensive validation of APOLLO system range accuracy has not been possible because of centimeter-scale deficiencies in computational models of the Earth-Moon range, and because APOLLO lacked an internal timing calibration system. Here, we report on the development of a system that enables in-situ calibration of the timing response of the APOLLO apparatus, simultaneous with lunar range measurements. The system was installed in August 2016. Preliminary results show that the APOLLO system can provide lunar range measurements with millimeter accuracy.

**Introduction**
There are manifold motivations for high-precision tests of gravity, including the incompatibility of General Relativity and Quantum Mechanics, and the search for the nature of dark energy and dark matter. Incisive tests of General Relativity are hard to come by, but Lunar Laser Ranging (LLR) can comment on a range of speculative gravitational theories through constraints on the strong and weak equivalence principles, Yukawa interactions and the time-rate-of-change of Newton's constant $G$ [1]. LLR is also sensitive to relativistic effects such as gravitomagnetism and geodetic precession [2]. Beyond gravitational physics, LLR can also constrain violations of Lorentz Invariance through searches for preferred directions in space [3], and dark-energy-inspired alternative gravity theories such as braneworld models [4]. A review of the physics reach of LLR is provided in Reference [1].

Since 2006, the Apache Point Observatory Lunar Laser-ranging Operation (APOLLO) has acquired millimeter-precision range measurements at the Apache Point Observatory 3.5m telescope in New Mexico, USA. Compared to other LLR stations, APOLLO benefits from a larger telescope aperture and better atmospheric conditions (the median seeing is 1.1 arcsecond), which result in 100-1000 times stronger return rates. There are five ranging targets on the lunar surface: three arrays deployed by Apollo missions (Apollo 11, 14 and 15), and two French-built arrays deployed on Russian rovers (Lunokhod 1 and 2). At present, about 50% of the APOLLO range measurements are made to the Apollo 15 array, 20% each to Apollo 11 and 14, and 5% each to Lunokhod 1 and 2 (see Figure 1).

**Lunar Ranging Apparatus**
A full description of the APOLLO apparatus is available in Reference [5]. Here, we provide a brief overview to help understand the implementation of the new timing calibration system. The laser used for lunar ranging emits at 532 nm (Nd:YAG, Q-switched and frequency doubled), with a 90 ps FWHM pulse width, a pulse energy of 115 mJ (~$10^{17}$ photons per pulse), and a 20 Hz repetition rate. The photon sensor is a 4x4 array of silicon avalanche photodiodes, fabricated by Lincoln Laboratory. Each pixel is 30 um in diameter, on a 100 um grid. A lenslet array is used to recover the fill-factor. The array images a 1.4x1.4 square-arcsecond patch of sky. The multi-pixel sensor allows APOLLO to process multiple photons per pulse, and also provides real-time tracking information during target acquisition and link optimization.

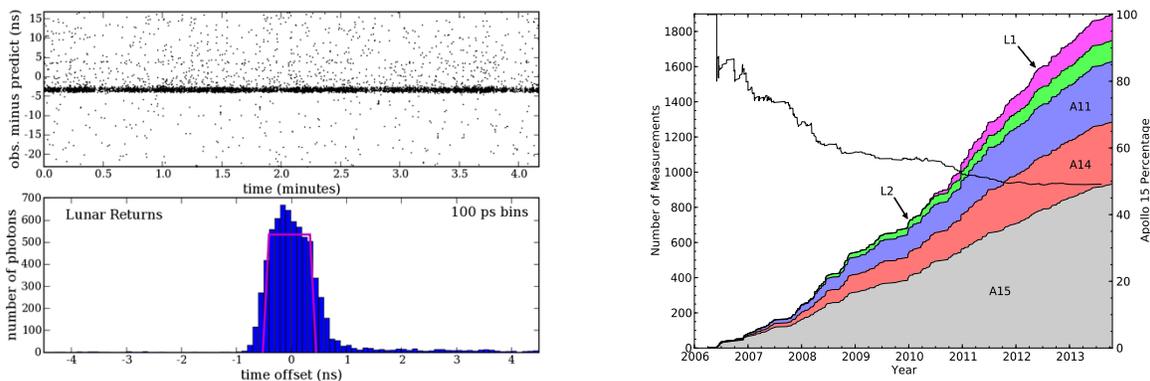

Figure 1: (Left) Top plot shows the observed minus predicted lunar range during a 4-minute observation on the Apollo 15 retroreflector array. Bottom plot shows a histogram of these data. The magenta line is a model of the time spread due to lunar libration. APOLLO collected 6,624 photons in 4 minutes, for a range precision of 0.8 mm. (Right) Cumulative distribution of APOLLO range measurements (normal points) to each lunar reflector, showing also that the reliance on Apollo 15 has gradually waned over time.

The range measurement is accomplished through a three-tiered timing system: a GPS-disciplined frequency standard (50 MHz), a system of counters to track the clock pulses (20 ns resolution), and a 12-bit time-to-digital converter (TDC) with 25 ps resolution and 100 ns range. The range time is defined as the difference in detection time between returns from an in-dome corner cube (the "fiducial") and the corner cube on the Moon (the "lunar").

**Checks on timing accuracy**

Until now, only weak, indirect tests on system timing accuracy have been available. For example, one can examine the residuals between range measurements and a model prediction (from a numerical planetary ephemeris). Such models are quite complex, only a few exist, and only one (the Planetary Ephemeris Program) is available for public use. The model that is believed to be the most complete, developed by the Jet Propulsion Laboratory, produces residuals with a weighted rms of ~1.5-2 cm. The PEP code residuals are approximately a factor of 2 worse. Although in principle one cannot conclude from the residuals alone whether the model or data is primarily to blame, there is strong evidence that points to model deficiencies (either in the form of missing terms in the range model, or errors in the implementation of the model, or both). To find and address these, the APOLLO collaboration has been working closely with the PEP team (Irwin Shapiro, Robert Reasenberg and John Chandler) on model refinements. In the meantime, however, the model-data comparison does not provide a useful validation of APOLLO data accuracy.

A test that can provide some confidence on APOLLO system timing accuracy is now described. Over very short time-scales (~1 hour), model-data residuals for a single reflector are expected to follow a linear trend. Indeed a study of APOLLO range measurements found this to be true, and, additionally, that the scatter in the residuals about the linear trend was consistent with the assigned range uncertainty. Furthermore, it was found that the linear trend could be accounted for by a modest modification of the telescope site coordinates in the range model [6]. While certainly not conclusive, this test provides additional confidence in the claim that APOLLO data accuracy is at the millimeter scale, and that, indeed, ephemeris models require further development to match that accuracy.

A more direct, conclusive test is certainly preferable. The rest of this document describes the design and implementation of a new apparatus, dubbed the Absolute Calibration System (ACS), to quantify the APOLLO system timing accuracy. Preliminary results from the ACS are also presented.

**Absolute Calibration System (ACS)**
In the early design stages for the ACS, the plan was to emulate the lunar ranging process, but using a controlled light source. In other words, a pair of optical pulses were to be generated on demand (to emulate the LLR launch and return signal). The time interval between pulses would be derived from a stable frequency standard (*e.g.* a cesium clock), and would be adjustable (in the neighborhood of the 2.5-second round-trip time to the Moon). The lack of optical sources at 532 nm with narrow pulse width (< 90 ps) and low jitter (< 50 ps) between electrical trigger and optical output forced a reconsideration of approach, and ultimately the implementation of a far more powerful scheme.

The final ACS design (shown in Figure 2) employs a fiber-cavity laser whose pulse repetition rate is controlled by adjusting the laser cavity length with a phase-locked-loop using a stable frequency standard (Microsemi 5071A cesium clock). In our system, the laser (Toptica Photonics PicoFYb 1064) produces a 80 MHz optical pulse train, with <10 ps width and <2 ps jitter relative to the frequency standard. Although this system does not allow the request of a pulse at a particular time, it ensures that the pulse-to-pulse separation is extremely well controlled (using the cesium frequency standard, pulses separated by 2 seconds have a timing jitter of a few picoseconds), thereby providing a timing calibration for the entire APOLLO timing system. During standard operation, the pulse train is held in a fully attenuated state. In response to the APOLLO ranging laser emission, a custom electro-optical system "slices out" a series of calibration laser pulses that coincide with the in-dome and lunar returns. By accepting calibration pulses during each of the fiducial and lunar return windows, an optical calibration ruler is overlaid atop the lunar range measurements. This calibrates exactly the quantity that matters most – the system timing response at the time of the range measurement.

**Preliminary Results with the ACS**
The ACS calibration results can be grouped into two categories: (1) independent studies of the relative stability of the GPS and cesium clocks, and (2) laser-based timing calibration, which exercises the full APOLLO timing apparatus. Each is addressed in turn.

The cesium clock was installed at APO in February 2016. The GPS-disciplined oscillator frequency is compared against the ACS cesium frequency standard using a Universal Counter (Agilent 53132A). The clock comparison is made whether APOLLO is actively ranging or not. Figure 3 (left) shows the instability of the GPS clock, expressed as a range error. Figure 3 (right) shows the clock-induced range error during nine lunar ranging sessions (nights). The nominal normal point uncertainties (the error bars in that plot) are a factor of two smaller than the scatter, consistent with an underestimation of the range error by the same factor. These results imply that, prior to ACS calibration, APOLLO reported errors that were a factor of two too small. Most importantly, however, the clock comparison data set makes it possible to correct for this clock-induced error. Furthermore, APOLLO has logged statistics of the GPS-disciplined system clock over its entire >10-year lifetime. Preliminary studies suggest that, using this data log, it will be possible to back-correct the archival APOLLO data, thereby reducing the clock-induced errors by a factor of two. Going forward, APOLLO will use the cesium clock as its main 50 MHz reference (in place of the GPS-disciplined clock), obviating this issue.

To close, results from an ACS calibration, taken simultaneously with lunar ranging to Apollo 15, are presented. Figure 4 shows two representations of data from the 500-second-long run. At left, the TDC bin for each detected photon as a function of (lunar) laser shot number. In this space, the ACS photons cluster in bands, and drift of the TDC value of each band indicates clock drift between the ACS cesium clock and the APOLLO system clock (GPS). At right are the same data, but with the lunar return time prediction subtracted (and plotted as a histogram), which clusters the lunar signal into a peak. Even though the ACS and lunar photons are overlaid in the TDC space, precise knowledge of the relative phasing between the cesium and GPS clocks allows for accurate tagging of these two populations, though some photons (shown in blue) cannot be confidently tagged, and are omitted from the following analysis.

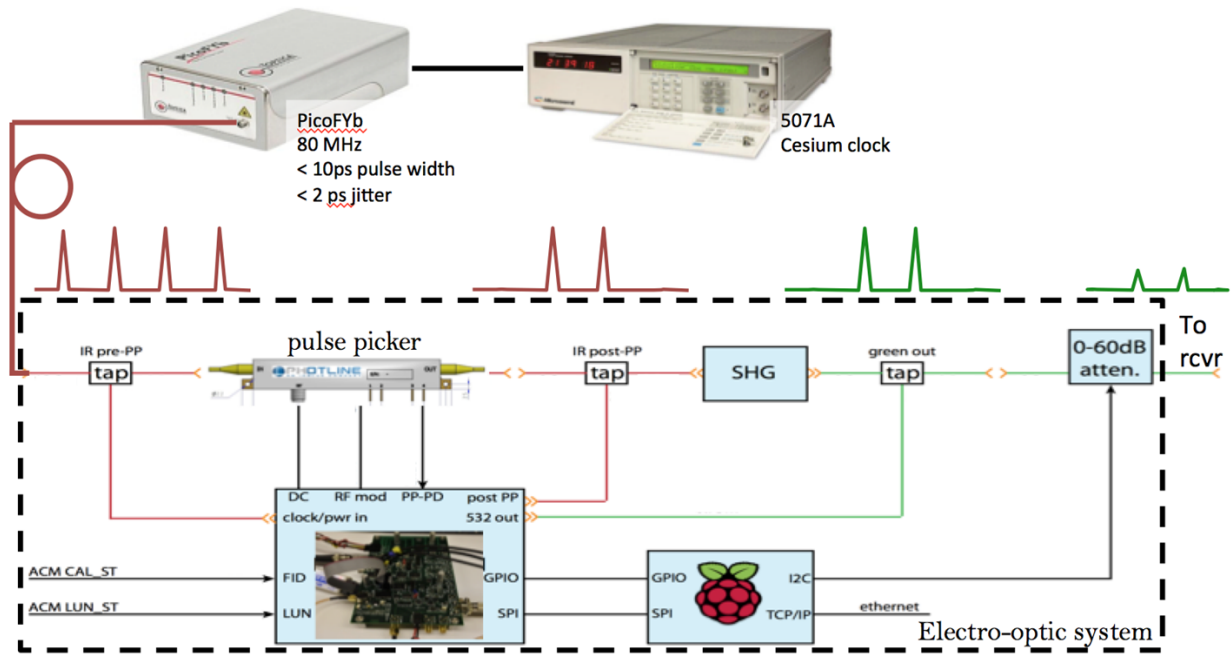

Figure 2: Schematic of the ACS system, including a block diagram of the electro-optical system that slices pulses out of the optical pulse train. At top left is the PicoFYb laser, producing an 80 MHz pulse train, with < 10 ps pulse width and < 2 ps jitter relative to the frequency standard. The cesium frequency standard (Microsemi 5071A) is shown at top middle. To select pulses out of the train, an electro-optical modulator (Mach-Zender) is usually held in an opaque state except when a pulse sequence is requested. The pulse request signal is generated by a custom circuit board that generates a pulse transmission request based on inputs related to the lunar laser fire signals from the APOLLO system. The pulse transmission request window placement and size (number of transmitted pulses) are both remotely adjustable via a Raspberry Pi. Following the pulse slicing, the optical signal is frequency doubled by a Second Harmonic Generator (SHG), and then attenuated by a remote-controlled attenuation device such that 1 photon per pulse, on average, is delivered to the APD array.

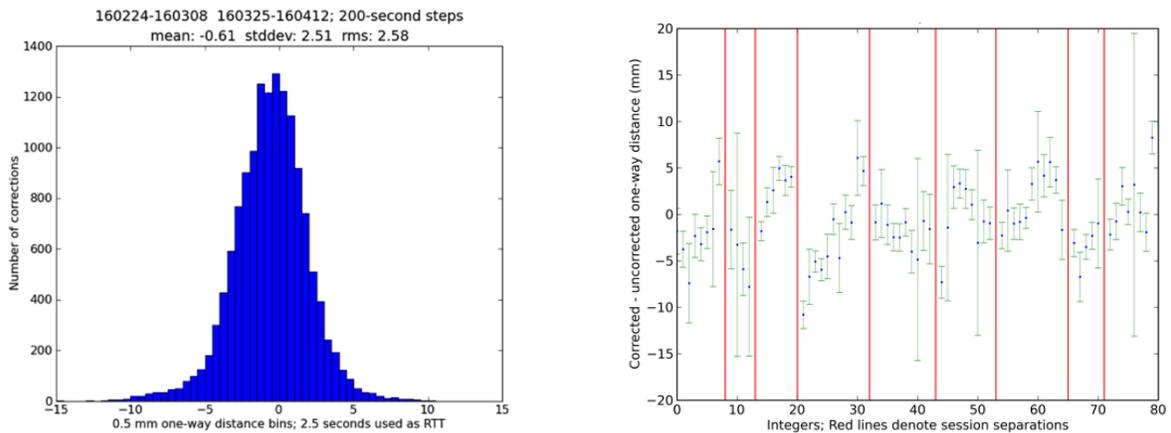

Figure 3: (Left): Distribution of the frequency differences between the GPS-disciplined clock and the cesium clocks, expressed as a range error during a 2.5 second round-trip measurement. The standard deviation of the distribution, which quantifies APOLLO's range measurement accuracy, is 2.5 mm. This clock-error-induced range uncertainty can be corrected. (Right): Range errors computed from clock comparison during lunar ranging sessions. Red lines demarcate different observing sessions (nights). Error bars are the nominal (statistical) normal point uncertainty. The range error is observed to exceed the uncertainties by approximately a factor of two, but can be corrected using the clock-comparison data.

These data reveal much good news about APOLLO system timing. One useful test, for example, compares the measured time difference between an ACS photon in a fiducial gate and an ACS photon in the corresponding lunar return gate (roughly 2.5s later), against integer multiples of 12.5000 ns. The resulting distribution of timing errors is Gaussian, with range offset of ~0.5 mm (the exact value depends on how data is combined across all of the APD channels). On longer time-scales (~1000 shots = 50 seconds), the ACS calibration reveals (and can be used to correct for) clear evidence for clock drift, with a fractional frequency offset of $-4.3 \times 10^{-12}$ during shots 1-7000, and $+4 \times 10^{-12}$ during shots 7000-9000 (the corresponding round-trip range errors for a 2.5 s travel time are -1.6 mm and 1.5 mm). Furthermore, the ACS can be used to calibrate the timing offsets between the different APD channels. This was previously done with lunar range data compiled over months, but can now be done with minutes of ACS data, and is found to be more accurate (as measured by the smaller width of the composite signal of the co-added channels). The full power of the ACS has not yet been realized – new system calibration tests are under development. But even with this first 500-second observation with ACS+LLR, we see no evidence for substantial timing inaccuracies in APOLLO. Now that the millimeter-scale measurement errors have been identified, the APOLLO collaboration can set out to eliminate them.

In conclusion, a new APOLLO calibration system has been constructed to measure timing accuracy. Preliminary results from calibration measurements, both independent of, and synchronous with lunar range measurements reveal that the APOLLO range accuracy is at the millimeter scale. Going forward, the more stable cesium frequency standard can replace the existing GPS-disciplined APOLLO system clock, thereby removing some sources of timing uncertainty. With future ACS measurements, the remaining timing systematics can be tracked down and mitigated.

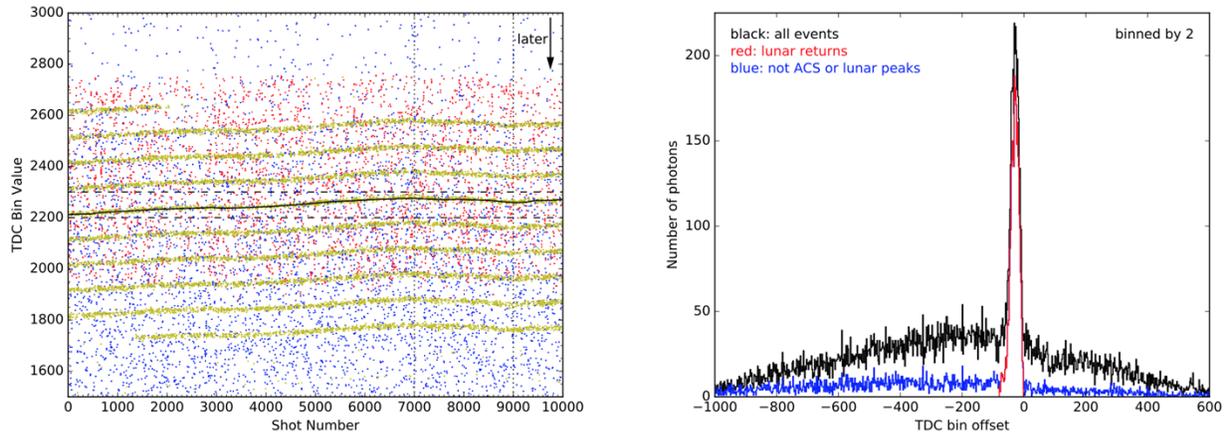

Figure 4 (Left): raw data from an LLR+ACS run obtained over 500 s on 2016-09-12. Each dot represents a photon detection during the lunar gate. TDC time measurements are 25 ps per bin; later photons appear lower in the plot. Yellow dots have been identified as ACS pulses based on phase relative to the APOLLO clock (GPS). Red dots indicate photons tagged as lunar returns. Blue dots represent the remainder—not tagged as lunar or ACS—and largely represent background, slow avalanches due to diffusion, or delayed crosstalk events in the APD. The solid black line is constructed from an independent measurement (using the Universal Counter) of the APOLLO clock frequency referenced to the cesium clock. Right: histogram of the lunar-prediction-corrected TDC values, showing the high visibility of the lunar signal—even though weaker than the aggregate ACS signal and spread over approximately the same TDC region. The gap between black and blue traces is attributable to the yellow dots in the left panel, smeared by the lunar prediction adjustments. The masking by ACS photons in the left panel looks five times worse here than it really is, because only one in five stripes is "live" for a given shot.